\documentclass{iopart}

\usepackage{latexsym}
\usepackage{hyperref}
\usepackage{longtable}
\usepackage{array}
\usepackage{dcolumn}
\usepackage{epsfig}
\usepackage{amsfonts}
\usepackage{bm}
\usepackage{graphicx}

\begin{document}

\title{Directional magnetoelectric effects in MnWO$_4$: Magnetic sources of the electric polarization}

\author{P. Tol\'{e}dano$^1$, B. Mettout$^1$, W. Schranz$^2$ and G. Krexner$^2$}

\address{$^{1}$Laboratory of Physics of Complex Systems, University of Picardie, 33 rue Saint-Leu, 80000 Amiens, France}
\address{$^{2}$Faculty of Physics, University of Vienna, Boltzmanngasse 5, A-1090 Vienna, Austria}

\ead{pierre.toledano@wanadoo.fr}

\begin{abstract}
The ferroelectric order and magnetic field induced effects observed in the spiral phase of MnWO$_4$ are described theoretically.
It is demonstrated explicitly that the Dzyaloshinskii-Moriya antisymmetric interactions contribute to the correlation between spins and electric dipoles in the incommensurate and commensurate ferroelectric phases of magnetic multiferroics. However, other single-site symmetric interactions are shown to be involved in the magnetoelectric process, suggesting the possible existence of an electric polarization originating from purely symmetric effects.
\end{abstract}

\section{Introduction}
 A typical feature of multiferroic materials undergoing a transition to an elliptic spiral ferroelectric phase  \cite{Fiebig,Kimura_Review,Cheong}, is the existence of spectacular magnetoelectric effects, such as the polarization flops observed in $TbMnO_{3}$ \cite{Kimura_Nature} and $DyMnO_{3}$ \cite{Strempfer}, or the sign reversal of $P_{y}$ disclosed under magnetic field in $TbMn_{2}O_{5}$  \cite{Hur}.
 The theoretical description of these effects requires knowledge of the order-parameter symmetry
 associated with the ferroelectric transition. Besides, the orientation of the applied magnetic field with respect to the magnetic spins influences the stability range of the spiral phase and the polarization flop process. This property was recently illustrated by remarkable magnetic field induced effects observed in the ferroelectric phase of $MnWO_{4}$:
The stability range of the phase depends on the direction of the applied field \cite{Lauten,Taniguchi,Sagayama,Taniguchi2008a}, which induces a high-field polarization-flop transition \cite{Taniguchi2008b}. Here we show that these directional effects result from the order-parameter symmetry associated with the ferroelectric phase of MnWO$_4$, the stability of which depends on the respective orientations of the magnetic field and magnetic easy axis.\\
It is of fundamental interest to understand what is the microscopic mechanism behind the magnetoelectric coupling in multiferroics. It is generally believed \cite{Cheong}, that in RMnO$_3$ (R=Tb, Dy, Gd) the \textit{antisymmetric} Dzyaloshinskii-Moriya (DM) interaction is the microscopic origin for the ferroelectric polarization.
For RMn$_2$O$_5$ the spins are almost collinear in the main ferroelectric phase and it is claimed \cite{Sushkov} that \textit{symmetric} exchange striction induces ferroelectricity.
To gain insight into the microscopic mechanism for ferroelectricity in MnWO$_4$ we express the order-parameters in terms of the magnetic spins and provide the correspondence between spins and electric dipoles.
This analysis confirms that the
Dzyaloshinskii-Moriya (DM) interactions \cite{Katsura,Sergienko} contribute to the
electric polarization in the incommensurate spiral phase but, additionally, shows that they participate as well in the formation of the dipole moments in the \textit{commensurate} ferroelectric phases of magnetic multiferroics. We also find that the DM interaction is not the only microscopic source of the polarization: Other \textit{symmetric} interactions are shown to be involved in the magnetoelectric process, suggesting the possible existence of a polarization induced by purely symmetric effects.\\

\section{P$\rightarrow$AF3$\rightarrow$AF2$\rightarrow$AF1 Transitions}
Below its wolframite-type paramagnetic (P) structure, of monoclinic $P2/c1'$ symmetry, MnWO$_4$ undergoes three successive magnetic phase transitions \cite{Lauten,Taniguchi} at 13.5 K ($T_{N}$), 12.7 K ($T_{2}$) and 7.6 K ($T_{1}$). They lead, respectively, to an incommensurate magnetic phase (AF3), an incommensurate elliptical spiral phase (AF2) displaying an electric polarization $\vec P // b$, and a commensurate magnetic phase (AF1). The transition wave-vectors \cite{Lauten} are $\textbf{k}_{inc} = (-0.214, \frac{1}{2}, 0.457)$ for AF2 and AF3, and $\textbf{k}_{com} =(\pm\frac{1}{4},\frac{1}{2},\frac{1}{2})$ for AF1. They are associated with two bidimensional irreducible corepresentations  of the $P2/c1'$ space group \cite{Kovalev}, denoted by $\Gamma^{\textbf{k}1}$ and $\Gamma^{\textbf{k}2}$ in \cite{Lauten}, whose generators are given in Table I.

\begin{table}[h]
\caption[TABLE I:]{Generators of the irreducible corepresentations $\Gamma^{\textbf{k}1}$ and $\Gamma^{\textbf{k}2}$ of the  $P2/c1'$ paramagnetic space group, deduced from the irreducible representations $\hat \tau_1$ and $\hat \tau_2$ of $G_k=m_y$, given in Kovalev´s tables \cite{Kovalev} for ${\bf k}_{inc}$ and ${\bf k}_{com}$ . T is the time reversal symmetry. For AF2 and AF3, $\varepsilon \approx 0.457 \pi$, $\varepsilon_1 \approx 0.418 \pi$ and $\varepsilon_2=2\varepsilon$. For AF1, $\varepsilon=\varepsilon_1=\pi/2$ and $\varepsilon_2=\pi$.}
\setlength{\arraycolsep}{0.5mm}
\renewcommand{\arraystretch}{0.7}
$\begin{array}{cc|cccccc}
\hline
  P2/c1'  & & (\sigma_{y}|0 0 \frac{c}{2}) & (I|000) & T & (E|a00) & (E|0b0) & (E|00c) \\ \hline
\\
  \Gamma^{\textbf{k}1} & \begin{array}{c} {\bar S{_1}} \\ {\bar S{_1}^*} \end{array} &
               \left(
                 \begin{array}{cc}
                   e^{i\varepsilon} &    \\
                     & e^{-i\varepsilon} \\
                 \end{array}
               \right) &
               \left(
                 \begin{array}{cc}
                     & 1 \\
                   1 &   \\
                 \end{array}
               \right) &
                              \left(
                 \begin{array}{cc}
                   -1 &   \\
                     & -1 \\
                 \end{array}
               \right)  &
               \left(
                 \begin{array}{cc}
                   e^{i\varepsilon_1} &    \\
                     & e^{-i\varepsilon_1} \\
                 \end{array}
               \right) &
               \left(
                 \begin{array}{cc}
                      -1 & \\
                   & -1   \\
                 \end{array}
               \right)  &
               \left(
                 \begin{array}{cc}
                      e^{i\varepsilon_2} & \\
                  & e^{-i\varepsilon_2}     \\
                 \end{array}
               \right)  \\
   \Gamma^{\textbf{k}2} & \begin{array}{c} {\bar S{_2}}\\ {\bar S{_2}^*} \end{array} &
               \left(
                 \begin{array}{cc}
                   -e^{i\varepsilon} &    \\
                     & -e^{-i\varepsilon} \\
                 \end{array}
               \right) &
               \left(
                 \begin{array}{cc}
                     & 1 \\
                   1 &   \\
                 \end{array}
               \right) &
                             \left(
                 \begin{array}{cc}
                   -1 &   \\
                     & -1 \\
                 \end{array}
               \right)  &
               \left(
                 \begin{array}{cc}
                   e^{i\varepsilon_1} &    \\
                     & e^{-i\varepsilon_1} \\
                 \end{array}
               \right) &
               \left(
                 \begin{array}{cc}
                      -1 & \\
                   & -1   \\
                 \end{array}
               \right)  &
               \left(
                 \begin{array}{cc}
                      e^{i\varepsilon_2} & \\
                  & e^{-i\varepsilon_2}     \\
                 \end{array}
               \right)  \\
               \\ \hline
\end{array}$
\end{table}

The complex amplitudes  transforming according to  $\Gamma^{\textbf{k}1}$ and $\Gamma^{\textbf{k}2}$, which form the order-parameter components, are denoted by $\bar{S_{1}} = S_{1}e^{i \theta_{1}}$, $\bar{S_{1}}^{\ast} = S_{1}e^{-i \theta_{1}}$ and $\bar{S_{2}} = S_{2}e^{i \theta_{2}}$, $\bar{S_{2}}^{\ast} = S_{2}e^{-i \theta_{2}}$. For $\textbf{k} = \textbf{k}_{inc}$ the order-parameter invariants $\Im_{1} = S_{1}^{2}$, $\Im_{2} = S_{2}^{2}$ and $\Im_{3} = S_{1}^{2}S_{2}^{2}\cos(2\varphi)$, with $\varphi = \theta_{1}-\theta_{2}$, yield the Landau expansion:

\begin{eqnarray}\label{free-energy density}
    \Phi_{1}(T,S_{1},S_{2},\varphi) = \Phi_{10}(T) + \frac{\alpha_{1}}{2}S_1^2 + \frac{\beta_{1}}{4}S_1^{4} + \frac{\alpha_{2}}{2}S_2^{2} + \frac{\beta_{2}}{4}S_{2}^{4} + \nonumber\\
    +\frac{\gamma_{1}}{2}S_1^2S_2^2cos(2\varphi) +  \frac{\gamma_{2}}{4}S_1^4S_2^4cos^2(2\varphi) + \ldots
\end{eqnarray}

Minimizing $\Phi_{1}$ shows that five distinct stable states, denoted I-V, may arise below the P-phase for different equilibrium values of $S_{1}$,$S_{2}$ and $\varphi$, summarized in Fig.1(a). The theoretical phase diagram shown in Fig.1(b) gives the location of the phases in the $(\gamma_{1},\alpha_{1}-\alpha_{2} )$ plane.

\begin{figure}
\begin{center}
\includegraphics[scale=0.7]{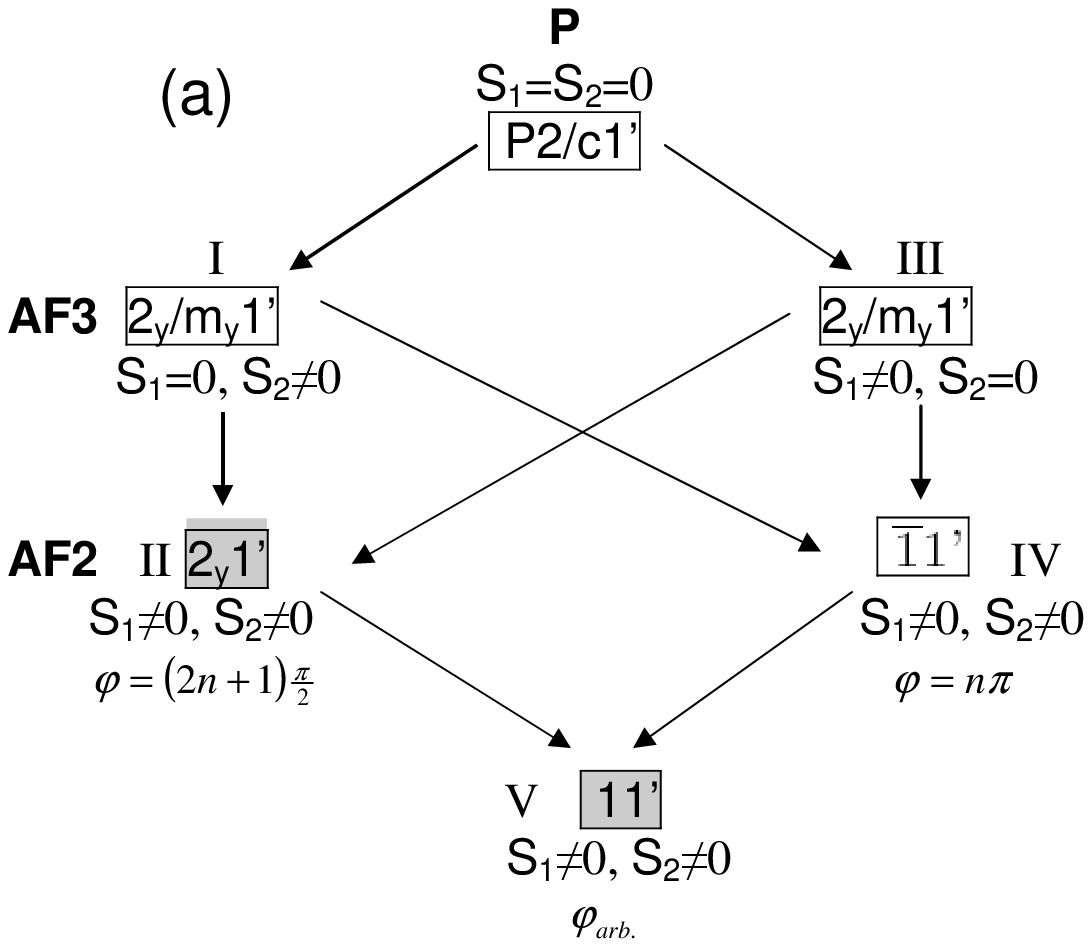}
\includegraphics[scale=0.45]{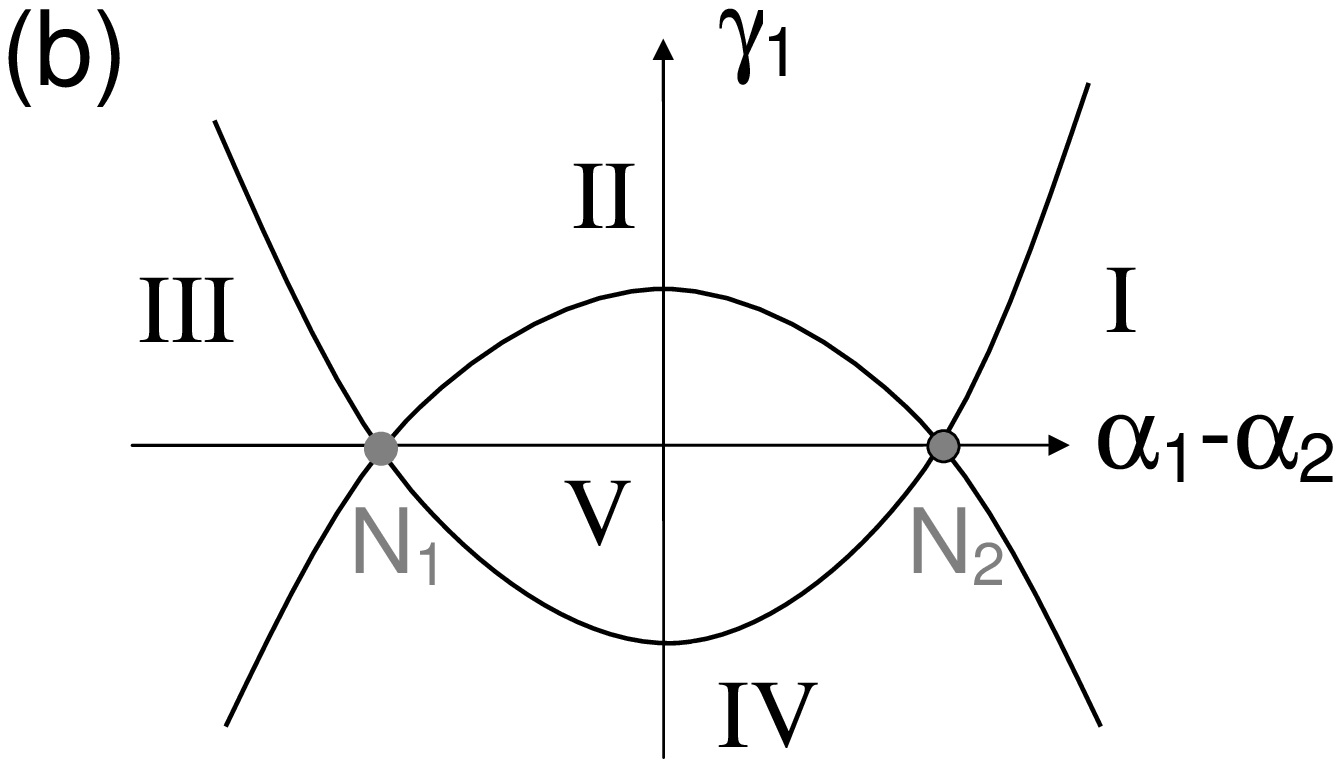}
\caption{(a) Connections between the magnetic point-groups of phases I-V induced by
 $\Gamma^{\textbf{k}1}+\Gamma^{\textbf{k}2}$ and equilibrium conditions fulfilled by the order-parameter in each phase. grey rectangles indicate ferroelectric phases.
(b) Phase diagram deduced from the minimization of $\Phi_1$ given in Eq.(\ref{free-energy density}), in the $(\gamma, \alpha_1-\alpha_2)$-plane. The phases are separated by second-order transition curves crossing at the four-phase points N$_1$ and N$_2$.}
\label{fig:1}
\end{center}
\end{figure}

Neutron diffraction data \cite{Lauten} show that the incommensurate magnetic AF3 phase induced by $\Gamma^{\textbf{k}2}$ corresponds to $S^e_1=0$ and $S^e_2 \neq 0$, coinciding with phase I in Fig.1(a). Its structure has the antiferromagnetic grey point-group $2_y/m_y1'$ , involving a doubling of the $b-$lattice parameter, and an incommensurate modulation of the spin density in the (x,z)-plane. The AF2 spiral phase, induced by
$\Gamma^{\textbf{k}1} + \Gamma^{\textbf{k}2}$ \cite{Lauten}, corresponds to phase II in Fig.1(a) with $S^e_{1}\neq 0$, $S^e_{2}\neq 0$, $\varphi = (2n+1)\frac{\pi}{2}$ and the magnetic symmetry $2_{y}1'$.
Adding the dielectric part of the free-energy density $\Phi_{1}^{D} = \delta P_{y}S_{1}S_{2}\sin\varphi + \frac{P_{y}^{2}}{2\varepsilon_{yy}^{0}}$ to Eq.(\ref{free-energy density}) yields the equilibrium polarization
\begin{equation}\label{equi_pol}
    P_{y}^{e} = \pm \delta \varepsilon_{yy}^{0}S_{1}^{e}S_{2}^{e}
\end{equation}

which changes its sign for opposite senses of the spiral configuration, as observed by Sagayama \textit{et al} \cite{Sagayama}.
The order-parameter $\bar{S_{2}}$, activated at the P$\rightarrow$AF3 transition, is frozen at the AF3$\rightarrow$AF2 transition, i.e. independent of temperature below T$_2$.
Therefore, Eq.(\ref{equi_pol}) expresses a \textit{linear} dependence of $P^e_y$ on $\bar{S^e_{1}}$, since $\delta \varepsilon_{yy}^{0}S_{1}^{e}$ acts as a temperature independent coupling coefficient.

In the AF2 and AF3 phases $\Phi_1$ can be truncated at the fourth degree since the eighth degree invariant is necessary only for stabilization of phase V (of Fig.1a).
Putting $\alpha_{1} = a_{1}(T-T_{0})$ and $\alpha_{2} = a_{2}(T-T_{N})$ and minimizing $\Phi_1$ of Eq.(\ref{free-energy density}) with respect to $S_1$ and $S_2$ one obtains the equilibrium values for the order-parameter components $S^e_1$ and $S^e_2$.\\
AF3 (T$_2<$T$<$T$_N$)corresponds to
\begin{equation}\label{equi_AF3}
   S^e_1=0 \quad \rm{and} \quad S^e_2=\pm \left[\frac{a_2}{\beta_2}(T_N-T)\right]^{1/2}
\end{equation}

 For T$\leq$T$_2$ (AF2 phase) one obtains
 \begin{equation} \label{equi_AF2}
 S_{1}^{e} = \pm\tilde{a}(T_{2}-T)^{1/2} \quad \rm{and} \quad S_{2}^{e} = \pm[\frac{\alpha_{2}}{\beta_{2}}(T_{N}-T_{2})]^{1/2}
 \end{equation}

 with
 \begin{equation} \label{parameters}
 \tilde{a} = (\frac{a_{2}\gamma_{1}+a_{1}\beta_{2}}{\beta_{1}\beta_{2}-\gamma_{1}^{2}})^{1/2} \quad \rm{and} \quad T_{2} = \frac{a_{2}\gamma_{1}T_{N}+a_{1}\beta_{2}T_{0}}{a_{2}\gamma_{1}+a_{1}\beta_{2}}
 \end{equation}

Inserting Eq.(\ref{equi_AF2})into Eq.(\ref{equi_pol}) one finds that the spontaneous polarization in the AF2 phase varies as

\begin{equation}\label{P(T)}
P_{y}^{e}(T) = \pm A(T_{2}-T)^{1/2}
\end{equation}

with $A = \delta \varepsilon_{yy}^{0}\tilde{a}[\frac{a_{2}}{\beta_{2}(T_{N}-T_{2})}]^{1/2}$. The dielectric permittivity follows a Curie-Weiss-type law around $T_{2}$ with

\begin{eqnarray}\label{eps(T)}
\varepsilon_{yy}(T) = \varepsilon_{yy}^{0}\left[1 - C\frac{T_{N}-T}{T-T_{2}}\right] \qquad
\rm{for} \quad T>T_{2}  \\ \nonumber
\varepsilon_{yy}(T) = \varepsilon_{yy}^{0}\left[1 - D\frac{T_{N}-T_{2}}{T_{2}-T}\right] \quad
\rm{for} \quad T<T_{2}
\end{eqnarray}

with $C = \frac{\delta^{2}a_{2}\varepsilon_{yy}^{0}}{a_{2}\gamma_{1}+a_{1}\beta_{2}}$ and $D = C\frac{\gamma_{1}^{2}-\beta_{1}\beta_{2}}{\gamma_{1}^{2}-4\beta_{1}\beta_{2}}$.

 \begin{figure}[h]
\begin{center}
\includegraphics[scale=0.35, angle=-90]{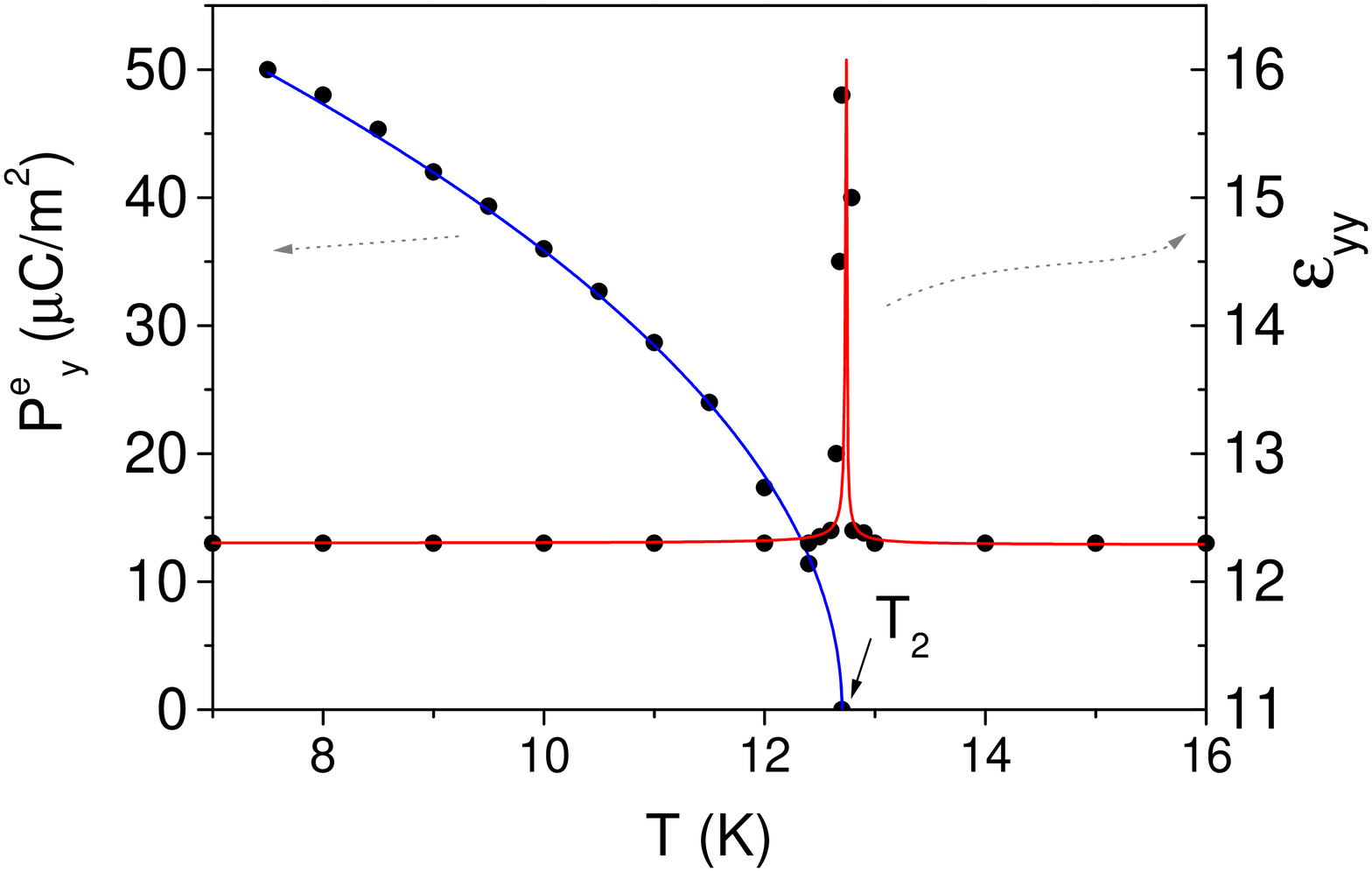}
\caption{Fit (lines) of the temperature dependence of the spontaneous polarization $P^e_y(T)$ and the dielectric permittivity $\varepsilon_{yy}(T)$ using Eq.(\ref{P(T)}) and Eq.(\ref{eps(T)}). The points are data from Taniguchi et al. \cite{Taniguchi}. The fit parameters are $A=21.8 \mu Cm^{-2}K^{-1/2}$, $C=0.001$, $D=0.0001$ and $\varepsilon_{yy}^0=12.3$.}
\label{fig:2}
\end{center}
\end{figure}

Fig.2 shows that $P_{y}^{e}(T)$ and $\varepsilon_{yy}(T)$ perfectly fit the experimental curves reported by Taniguchi et al. \cite{Taniguchi}.
This confirms the hybrid character of the ferroelectricity in spiral magnets, recently found \cite{Toledano2009,Tol_TbMn2O5} also for TbMnO$_3$ and TbMn$_2$O$_5$: The square-root temperature dependence of the polarization (Eq.(\ref{P(T)})) and the Curie-Weiss type behaviour of $\varepsilon_{yy}$ expressed by Eq.(\ref{eps(T)}) are typical for a \textit{proper} ferroelectric transition, whereas the low value of  $P_y \approx 40 \mu C/m^2$ measured at 10~K \cite{Kundys} is of the order found in \textit{improper} ferroelectrics.\\
Let us now turn to the AF1 phase.
Neutron diffraction results \cite{Lauten} indicate that the transition to the
commensurate AF1 phase triggers a \textit{decoupling} of $\bar{S_{1}}$ and $\bar{S_{2}}$.
Using Table 1 one finds that the lock-in at $T_{1}$, induced by $\Gamma^{\textbf{k}2}$, gives rise to the additional invariant $\Im_{4} = S_{2}^{4}\cos(4\theta_2)$.
The Landau expansion associated with the transition to AF1 is:

\begin{eqnarray}\label{free_2}
    \Phi_{2}(T,S_{2},\theta_{2}) = \Phi_{20}(T) + \frac{\alpha'}{2}S_{2}^{2} + \frac{\beta'}{4}S_{2}^{4} + \nonumber\\ + \frac{\gamma_{1}'}{4}S_{2}^{4}\cos(4\theta_{2})+\frac{\gamma_{2}'}{8}S_{2}^{8}\cos^{2}(4\theta_{2})
\end{eqnarray}

The equations of state show that three commensurate phases, denoted by I'-to-III', displaying a fourfold increased unit-cell (\textbf{b+c},\textbf{c-b},\textbf{2a+c}), may appear below $T_1$. The AF1 phase corresponds to phases I' or II', stable for $\cos(4\theta^e_2)=+1$ or $-1$, respectively, both described by the magnetic space group $C_a2/c$. Phase III', stable for $\cos(4\theta^e_2)=\frac{\gamma'_1 \beta'^2}{\gamma'_2 \alpha'^2}$, has the symmetry $C_ac$.

\section{Directional magnetic field (magnetoelectric) effects}
\begin{figure}[h]
\begin{center}
\includegraphics[scale=0.45, angle=0]{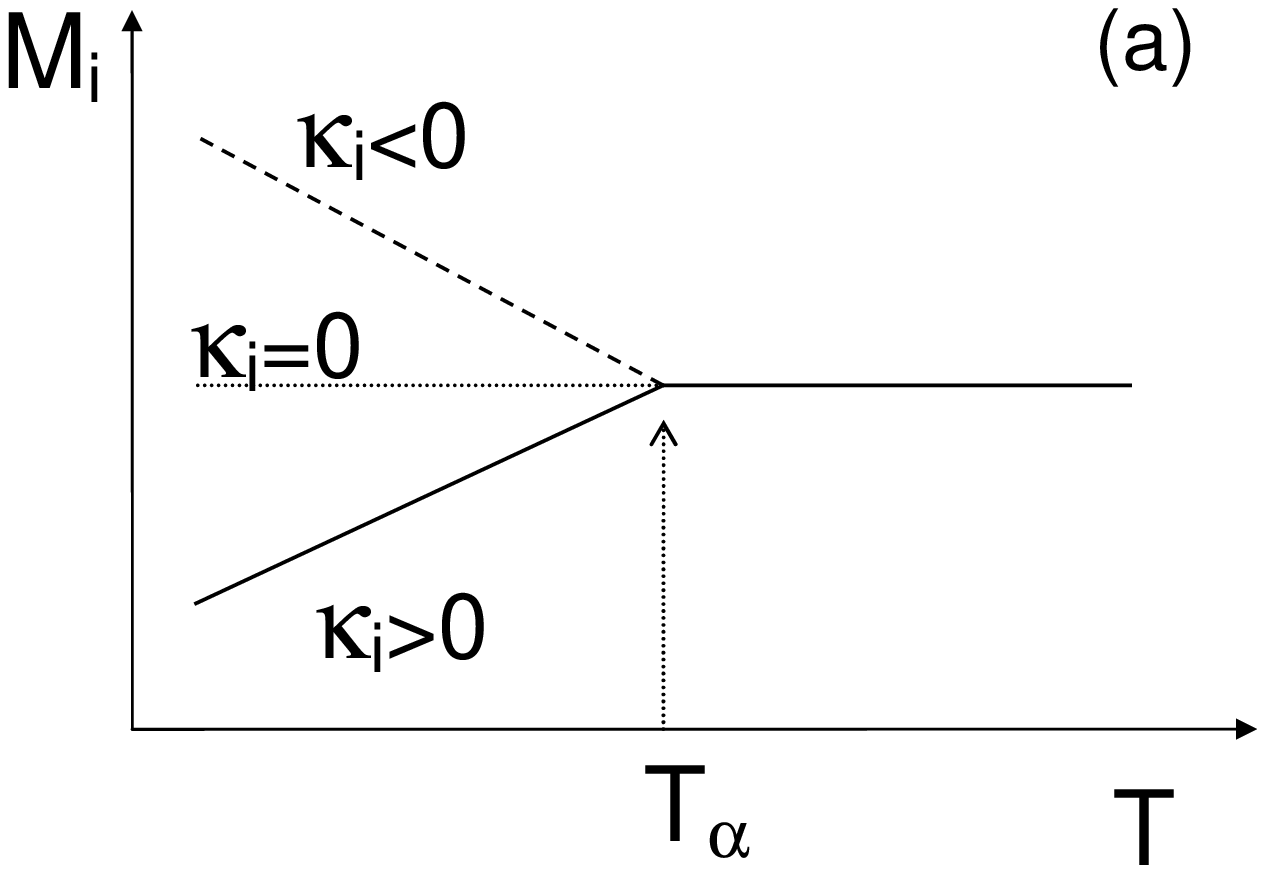}
\includegraphics[scale=0.45, angle=0]{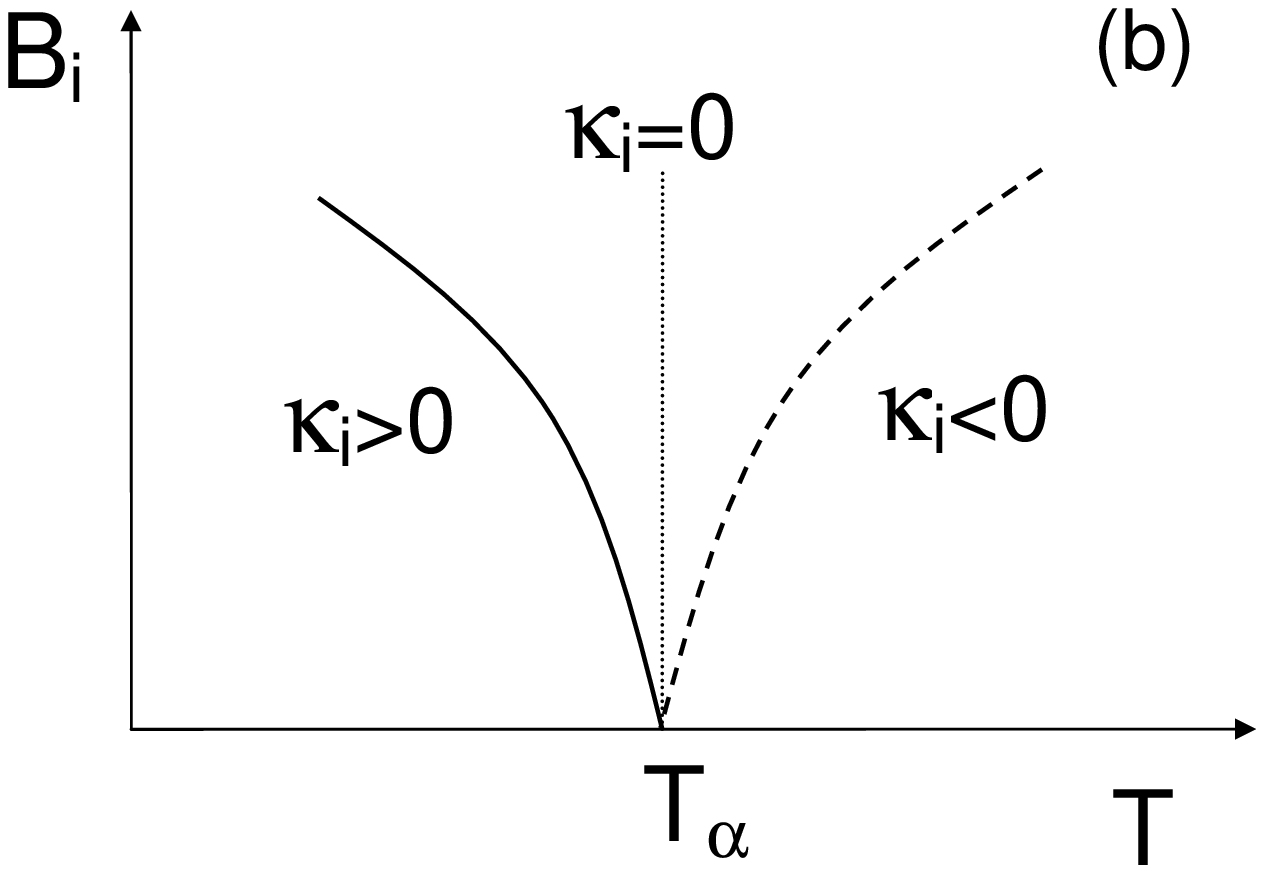}
\caption{Magnetizations M$_i$ (a) and magnetic field dependence (b) of transition temperatures calculated from Eqs.(\ref{Magnetization}) and (\ref{field-shift}).}
\label{fig:field dependence}
\end{center}
\end{figure}

The AF1 and AF2 order-parameters allow describing the magnetoelectric effects \cite{Taniguchi,Sagayama,Taniguchi2008a,Taniguchi2008b,Mitamura} observed in MnWO$_4$.
The magnetic phase diagram can be calculated by adding the magnetic part of the free-energy and the coupling invariants $\kappa_iM_i^2S_1^2 + \kappa'_iM_i^2S_2^2$

\begin{equation}\label{magnetic part}
\Phi_1^M=\frac{1}{2}\vec M \hat \mu \vec M - \vec B \vec M + \kappa_iM_i^2S_1^2 + \kappa'_iM_i^2S_2^2
\end{equation}

to the Landau-expansion Eq.(\ref{free-energy density}) or (\ref{free_2}).
$\hat \mu$ is the paramagnetic susceptibility tensor and i=x,y,z.\\
Due to the anisotropy of the magnetic free-energy the AF2 stability range depends on the angle
$\Psi_0=\frac{1}{2} tan^{-1} \left[ 2\mu_{xz}(\mu_{zz}-\mu_{xx})^{-1} \right]$ between $\vec B$ and the magnetic easy axis in the paramagnetic phase. If $\vec B$ is at an angle $\Psi$ with the x-axis, one finds:

\begin{equation}\label{T2ofPsi}
T_2(\Psi)-T_2=\varepsilon B^2 \{1-\frac{td^{-1}}{2}[t-\sqrt{t^2-4d}\cos2(\Psi-\Psi_0)]\}
\end{equation}

where $\varepsilon=(\beta_1 \kappa+\gamma_1 \kappa')(\beta_1 \alpha_1 + \gamma_1 \alpha_2)^{-1}$, $t=\mu_{xx}+\mu_{zz}$ and $d=\mu_{xx}\mu_{zz}-\mu_{xz}^2$. For $\varepsilon<0$ the AF2 stability range is maximum for $\vec B$ along the easy axis ($\Psi=\Psi_0$). It decreases when $\Psi$ increases from   $\Psi_0$ to $\Psi_0+\pi/2$, reducing to the stability range at zero field if $\mu_{xz},\mu_{xx}<<\mu_{zz}$. Such variation has been observed in the AF2 phase \cite{Taniguchi2008a,Arkenbout,Chaudhury}, in which $\Psi_0 \approx  35^o$ coincides with the direction of the spins in the (x.z)-plane.
When $\vec B$ is at an angle $\psi$ with the (x,z)-plane the AF2 stability range decreases when $\psi$ increases from $\psi=0$ to $\psi=\pi/2$, as reported experimentally \cite{Taniguchi2008a,Arkenbout}.\\
To get an overview about the magnetic field dependence of the various phases in MnWO$_4$ it may be useful to consider a simplified version of the free energy expansion by neglecting the non-diagonal parts of $\hat \mu$. Then by minimizing the free energy with respect to $M_i$ one obtains for the magnetization in x,y or z-direction

\begin{equation}\label{Magnetization}
M_i=\frac{B_i}{\mu_i+\kappa_iS^2}
\end{equation}

and the phase transition temperatures are shifted under applied magnetic field $B_i$ as

\begin{equation}\label{field-shift}
T_{\alpha}(B_i)=T_{\alpha}(0)-\frac{\kappa_i}{a \mu_i^2}B_i^2
\end{equation}

where $T_{\alpha}:=T_1,T_2$ or $T_N$ and $a>0$ is the bare expansion coefficient of the second order term in Eq.(\ref{free-energy density}). Eq.(\ref{field-shift}) shows, that the phase transition temperatures depend quadratically on the applied magnetic field and the sign of the magnetic field shift of T$_{\alpha}$ depends only on the sign of the coupling coefficient $\kappa_i$, which itself can be determined from the changes of $M_i$ with temperature (see Eq.(\ref{Magnetization})). Fig.\ref{fig:field dependence} sketches this behaviour.

Eqs.(\ref{Magnetization}) and (\ref{field-shift}) describe the magnetic phase diagram (see e.g. Fig.~5 of Ref\cite{Arkenbout}) perfectly. The strongest downshift of T$_1$ occurs in x-direction, which corresponds also to the strongest negative anomaly in M$_x$ at T$_1$, implying $\kappa_x>>\kappa_z>0$ as seen from Fig.~3 of Ref\cite{Arkenbout}. This figure also shows that $\kappa_y<0$ at T$_1$ (M$_y$ displays an upwards anomaly at T$_1$) and therefore T$_1(B_y)$ increases with applied magnetic field. In this way all particular features of the phase diagram can be reproduced.\\
In the following we will briefly discuss the effect of a very high magnetic field B$_y$.
Above a threshold field $B_y^{th}$, given by $\Phi_1(T_1,S^e_1,S^e_2)-B_y^{th}M_y^{II}=\Phi_2(T_1,S^e_2)-B_y^{th}M_y^{III'}$, the AF2 phase switches to the high-field phase III', which cancels $P_y$ and gives rise to the polarization:

\begin{equation}\label{polarizationIII'}
P_{x}^{III'} = - \delta' \varepsilon_{xx}^{0}S_{2}^{e2}sin(2\theta^e_2)
\end{equation}

deduced from the dielectric free-energy $\Phi_{2}^{D} = \delta' P_{x}S_{2}^2\sin(2\theta_2) + \frac{P_{x}^{2}}{2\varepsilon_{xx}^{0}}$. The onset of $P_x^{III'}$ occurs in correlation with the vanishing of $P^e_y$ at the first-order AF2$\rightarrow$III' flop-transition, consistent with the $P_y \rightarrow P_x$ polarization-flop \cite{Taniguchi} observed above 10 T. Decreasing $B_y$ below $B^{th}_y$ switches back phase III' to AF2. If $B_y$ is canted at an angle $\phi$ with respect to the $b$-axis in the (x,z)-plane, the magnetoelectric coupling $\nu P_xM_xM_yS_1S_2\sin\varphi$ induces the polarization

\begin{equation}\label{cantedpolarization}
P_{x}^{e} = - \varepsilon^0_{yy}\nu \mu_{yy}^{-2}B_y^2S^e_1S^e_2\sin\varphi^e\sin2\phi
\end{equation}

Canting oppositely $B_y$ from the $b$-axis ($\phi \rightarrow -\phi$) \textit{reverses} $P^e_x$. Increasing again $B_y$ above $B^{th}_y$ yields an \textit{opposite sign} for $P_x^{III'}$ in phase III', as observed by Taniguchi \textit{et al} \cite{Taniguchi2008b}.

\section{Relating the order-parameters to magnetic spins}
To gain insight into the nature of the microscopic interactions, let us express $\bar{S_{1}}$ and $\bar{S_{2}}$ as a function of the magnetic spins in the commensurate phases I'-III'. Denoting $\bf{s_1}-\bf{s_8}$ the spins associated with the eight Mn$^{2+}$-ions of the corresponding fourfold primitive monoclinic unit-cell, one can write ${\bf s_i}=s_i^a \vec a + s_i^b \vec b + s_i^c \vec c$ (i=1-8), where $\vec a$, $\vec b$ and $\vec c$ are lattice vectors.
Projecting on $\Gamma^{\textbf{k}1}$ and $\Gamma^{\textbf{k}2}$ the matrices transforming the $s_i^{a,b,c}$-components gives:

\begin{eqnarray}\label{opspinrelation}
\bar{S_{1}}^{a,c}=L_1^{a,c}+iL_3^{a,c} \qquad \bar{S_{1}}^{b}=-L_4^{b}+iL_2^{b} \nonumber  \\
\bar{S_{2}}^{a,c}=-L_4^{a,c}+iL_2^{a,c} \qquad \bar{S_{2}}^{b}=L_1^{b}+iL_3^{b}
\end{eqnarray}

where the $\bar{S_{i}}^m$ (i=1,2 m=a,b,c) represent different forms of $\bar{S_{1}}$ and $\bar{S_{2}}$. The $L_i^m$ (i=1-4) are projections of the vectors

\begin{eqnarray}\label{projections}
{\bf L_1}={\bf s_1}-{\bf s_2}-{\bf s_7}+{\bf s_8}\\ \nonumber
{\bf L_2}={\bf s_1}-{\bf s_2}+{\bf s_7}-{\bf s_8}\\ \nonumber
{\bf L_3}={\bf s_3}-{\bf s_4}+{\bf s_5}-{\bf s_6}\\ \nonumber
{\bf L_4}={\bf s_3}-{\bf s_4}-{\bf s_5}+{\bf s_6}
\end{eqnarray}

In phase I'($\bar{S_{1}}^{a,b,c}=0, \cos4\theta_2^{a,b,c}=1$) one has $s_1^{a,c}=s_2^{a,c}=s_7^{a,c}=s_8^{a,c}=0$, $s_3^{b}=s_4^{b}=s_5^{b}=s_6^{b}=0$, $s_3^{a,c}=-s_4^{a,c}=-s_5^{a,c}=s_6^{a,c}$  and $s_1^{b}=-s_2^{b}=-s_7^{b}=s_8^{b}$.
It gives $\bar{S_{2}}^{a,c}=-L_4^{a,c}$ and $\bar{S_{2}}^{b}=L_1^{b}$.\\
In phase II'($\bar{S_{1}}^{a,b,c}=0, \cos4\theta_2^{a,b,c}=-1$)  $s_1^{a,c}=-s_2^{a,c}=s_7^{a,c}=-s_8^{a,c}=-s_3^{a,c}=s_4^{a,c}=s_5^{a,c}=-s_6^{a,c}$ and $s_1^{b}=-s_2^{b}=-s_7^{b}=s_8^{b}=s_3^{b}=-s_4^{b}=s_5^{b}=-s_6^{b}$.
It yields $\bar{S_{2}}^{a,c}=-L_4^{a,c}+iL_2^{a,c}$ and $\bar{S_{2}}^{b}=L_1^{b}+iL_3^{b}$.

\begin{figure}[h]
\begin{center}
\includegraphics[scale=0.6]{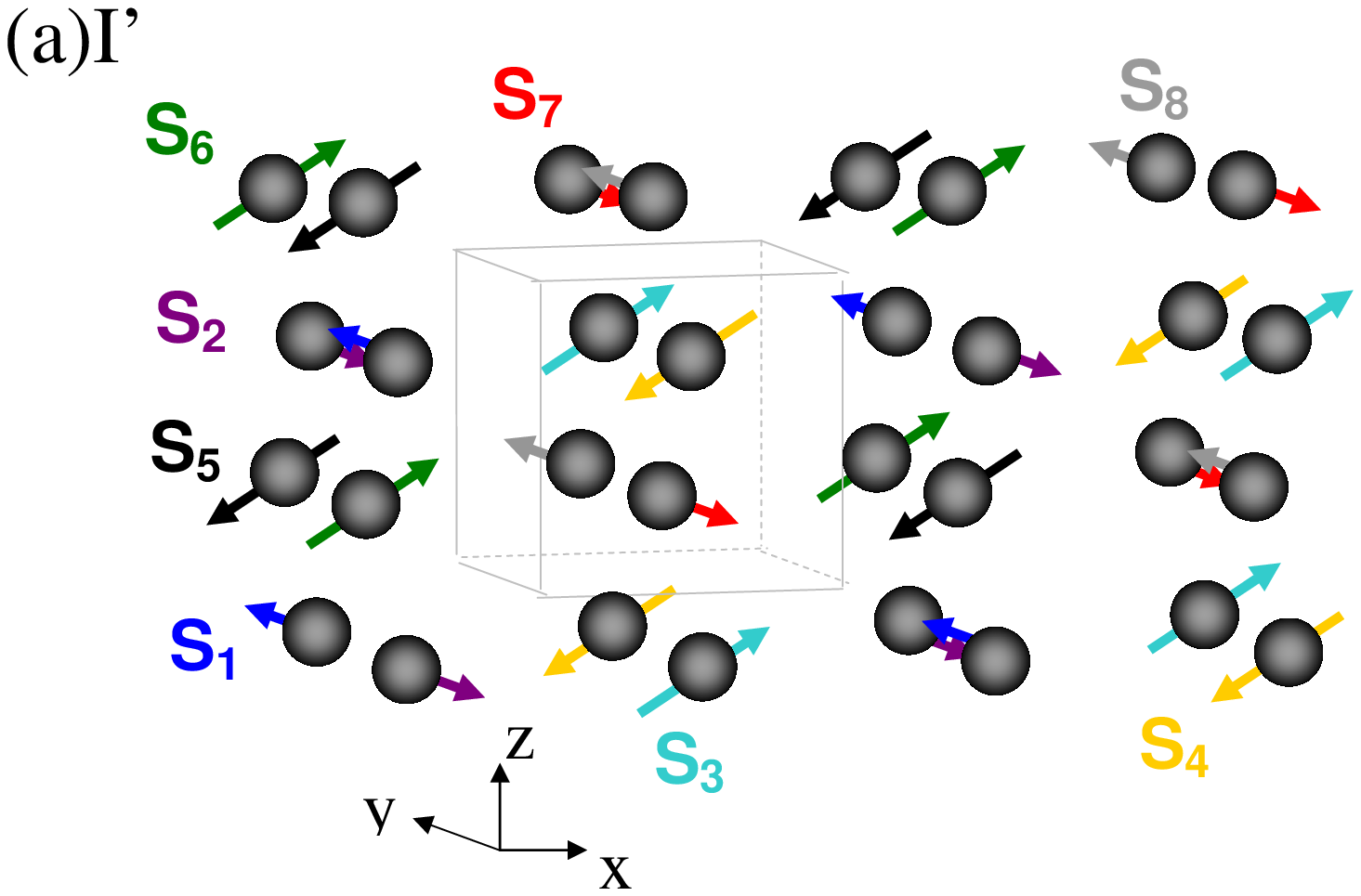}
\includegraphics[scale=0.6]{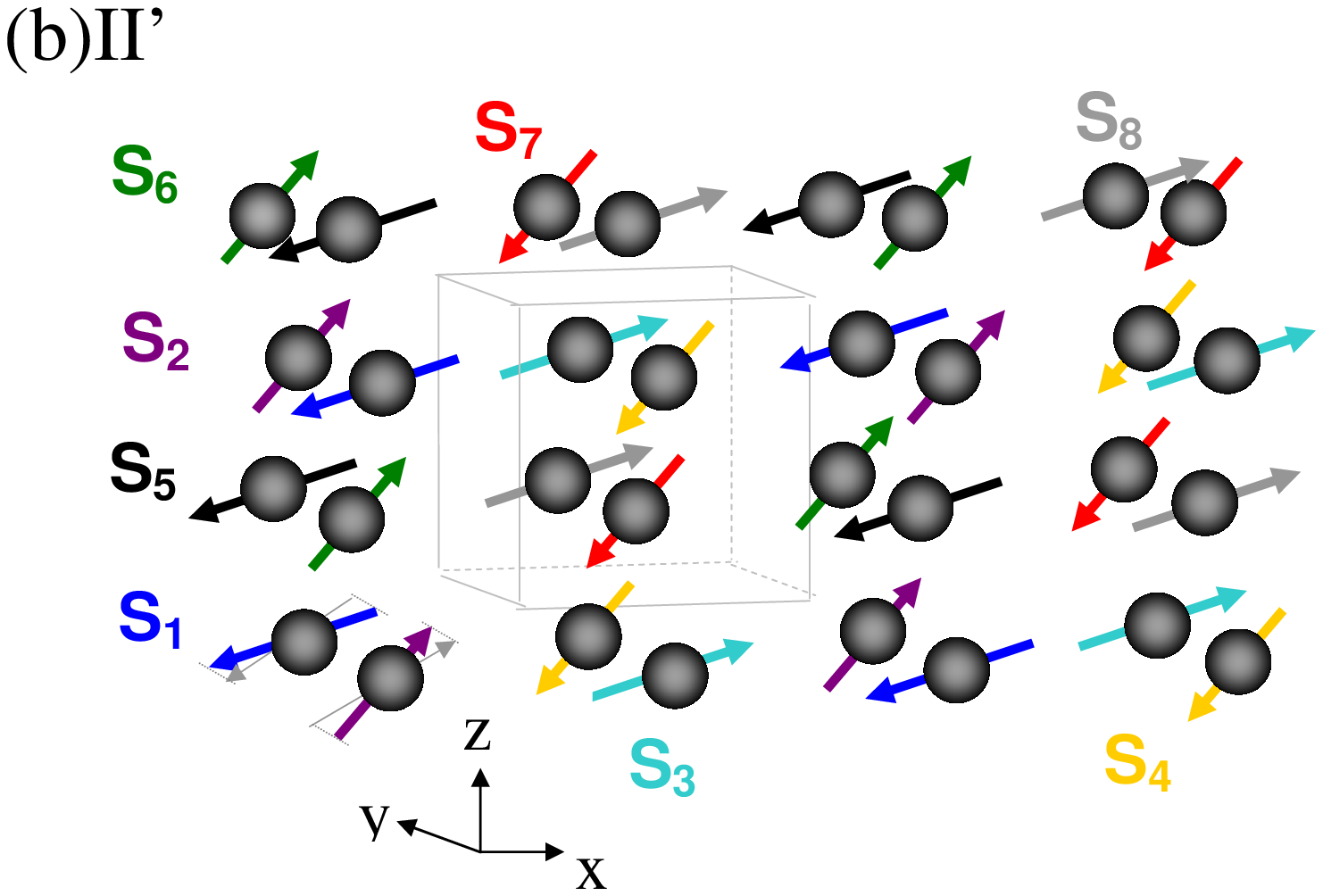}
\caption{Calculated magnetic structures of (a) phase I' and (b) phase II', described in the text.
The two additional grey arrows oriented in (x,z)-plane located at the two atoms in the bottom left of (b) indicate the previously refined\cite{Lauten} magnetic structure of the AF1-phase. The
coordinates of the eight Mn$^{2+}$ ions corresponding to the spins ${\bf s}_1-{\bf s}_8$, in the primitive monoclinic unit-cell of the two structures are for ion 1$\left(\frac{1}{2},y-1,\frac{5}{4}\right)$,
2$\left(\frac{1}{2},y,\frac{5}{4}\right)$,
3$\left(\frac{3}{2},y,\frac{5}{4}\right)$,
4$\left(\frac{3}{2},y,\frac{9}{4}\right)$,
5$\left(\frac{1}{2},{\bar y},\frac{3}{4}\right)$,
6$\left(\frac{1}{2},{\bar y},\frac{7}{4}\right)$,
7$\left(\frac{3}{2},{\bar y},\frac{7}{4}\right)$,
8$\left(\frac{3}{2},1-y,\frac{7}{4}\right)$.}
\label{fig:4}
\end{center}
\end{figure}

Figs.~4(a) and 4(b) show that the magnetic structure of the AF1 phase of MnWO$_4$ \textit{coincides with the antiferromagnetic order of phase II'} (Fig.~4(b)), since the cancellation of the (a,c)-spin-components on the sites 1,2,7 and 8 in phase I' (Fig.~4(a)) was not observed \cite{Lauten} in AF1. The lack of  spin components along $\vec b$ reported in this phase \cite{Lauten} may be due to their relativistic origin, as suggested in \cite{Lauten}. In phase III'($\bar{S_{1}}^{a,b,c}=0, \theta_2^{a,b,c} \neq 0, \pi/4$), identified as the high-field phase of MnWO$_4$, the equilibrium spins are
$s_1^{a,c}=-s_2^{a,c}=s_7^{a,c}=-s_8^{a,c}$, $s_3^{a,c}=-s_4^{a,c}=-s_5^{a,c}=s_6^{a,c}$,
$s_1^{b}=-s_2^{b}=s_7^{b}=-s_8^{b}$ and $s_3^{b}=-s_4^{b}=s_5^{b}=-s_6^{b}$, with
$\bar{S_{2}}^{a,c}=-L_4^{a,c}+iL_2^{a,c}$ and $\bar{S_{2}}^{b}=iL_3^{b}$. \\
Other commensurate structures can take place for ${\bf k} = {\bf k}_{com}$ when the spin components associated with $\bar{S_{1}}$ and $\bar{S_{2}}$ order simultaneously. Two among these structures, denoted IV' ($\theta^e_1=0,\theta^e_2=\pi/2$) and V' ($\theta^e_1=\pi/2,\theta^e_2=0$), display the symmetry $2_y$ and the same form of $P_y \propto \frac{1}{2i}(\bar{S_{1}}\ \bar{S_{2}}^*- \bar{S_{2}}\ \bar{S_{1}}^*) = S_1 S_2 \sin\varphi$ as in the AF2 phase. Although these structures are not stabilized in MnWO$_4$, they are lock-in limits of the incommensurate AF2 structure, and can be used for investigating the microscopic origin of ferroelectricity in the spiral magnetic structure.
Since $\bar{S_{1}}$ and $\bar{S_{2}}$ are both realized by three independent combinations of spin components and taking into account the
equilibrium relationships ${\bf s}_1=-{\bf s}_2$, ${\bf s}_3=-{\bf s}_4$, ${\bf s}_5=-{\bf s}_6$ and ${\bf s}_7=-{\bf s}_8$ which hold for all commensurate structures of MnWO$_4$, $P_y$ is expressed as the sum of nine terms:
\begin{eqnarray} \label{Pyfrommagneticspins}
P_y=\delta_1 (s_1^{a2}+s_3^{a2}-s_5^{a2}-s_7^{a2}) +
\delta_2 (s_1^{b2}+s_3^{b2}-s_5^{b2}-s_7^{b2}) + \nonumber \\
\delta_3 (s_1^{c2}+s_3^{c2}-s_5^{c2}-s_7^{c2})
+\delta_4 (s_1^{a}s_1^{c}+s_3^{a}s_3^{c}-s_5^{a}s_5^{c}-s_7^{a}s_7^{c}) +  \nonumber \\
\delta_5 (s_1^bs_5^a-s_1^as_5^b+s_3^bs_7^a-s_3^as_7^b) +
\delta_6(s_1^bs_5^c-s_1^cs_5^b+s_3^bs_7^c-s_3^cs_7^b) +  \nonumber\\ \delta_7(s_1^bs_3^a-s_1^as_3^b+s_5^bs_7^a-s_5^as_7^b) +
\delta_8(s_1^bs_3^c-s_1^cs_3^b+s_5^bs_7^c-s_5^cs_7^b) + \nonumber\\ \delta_9(s_3^as_5^c-s_3^cs_5^a+s_1^cs_7^a-s_1^as_7^c)
\end{eqnarray}
The $\delta_1-\delta_4$ terms in Eq.(\ref{Pyfrommagneticspins}) are symmetric invariants involving a single atom. Their origin is entropic and due to on-site interactions. The $\delta_5-\delta_9$ terms represent Dzyaloshinskii-Moriya (DM) antisymmetric coupling interactions between neighbouring pairs of spins ${\bf s}_1,{\bf s}_5$ (or the equivalent pair ${\bf s}_3,{\bf s}_7$), ${\bf s}_1,{\bf s}_3$ (${\bf s}_5,{\bf s}_7$) and ${\bf s}_3,{\bf s}_5$  (${\bf s}_1,{\bf s}_7$). The DM interaction \cite{Dzyaloshinskii,Moriya} is currently assumed as the microscopic source of the polarization in the spiral structure of magnetic multiferroics \cite{Taniguchi2008b,Katsura}. Our results confirm explicitly this view in MnWO$_4$, but show that other \textit{symmetric} effects are also involved in the formation of the electric dipoles. Furthermore, equilibrium relationships of the spin components in phases IV' ($s_1^b=s_7^b=0,s_3^{a,c}=s_5^{a,c}=0$) and V' ($s_3^b=s_5^b=0,s_1^{a,c}=s_7^{a,c}=0$) preserve the symmetric and DM contributions in Eq.(\ref{Pyfrommagneticspins}). This indicates that the interactions giving rise to the polarization in the commensurate ferroelectric phases of multiferroic compounds are of the same nature than in the spiral phases.
The effect of the incommensurability in the x-z plane of the AF2 phase should result in further averaging of the spin densities without modifying essentially the invariants in Eq.(\ref{Pyfrommagneticspins}).
One should note that the DM interactions, as well as the Katsura-type contribution \cite{Katsura} ${\bf e}_{ij}\times{\bf s}_i\times{\bf s}_j$ (where ${\bf e}_{ij}$ is the distance vector joining neighboring Mn atoms i and j), cancel in Eq.(\ref{Pyfrommagneticspins}) when all spin components are cancelled except $s_1^a$ and $s_7^a$ in phase IV', or $s_3^a$ and $s_5^a$ in phase V',  $P_y$ keeping a finite value $P_y=\delta_1(s_1^{a2}-s_7^{a2})$ and  $P_y=\delta_1(s_3^{a2}-s_5^{a2})$ in phases IV' and V', respectively. Therefore symmetry considerations predict the existence of a polarization induced by interactions being neither of Dzyaloshinskii-Moriya nor of Katsura-type, although such effects may encounter considerable restrictions at the microscopic level.\\

\section{Summary and conclusion}
In summary, the present work clarifies the nature of the  ferroelectric order occurring in the spiral phase of MnWO$_4$, and gives a theoretical description of the field-induced effects observed in this compound. It confirms that the antisymmetric Dzyaloshinskii-Moriya interaction is involved in the formation of dipolar moments in the incommensurate and commensurate ferroelectric structures of magnetic multiferroics. It also shows that other symmetric effects participate in the ferroelectricity observed in these compounds, suggesting the possible existence of an unconventional ferroelectricity in magnetoelectric materials originating from purely symmetric interactions. Symmetric exchange interactions may mediate magnetoelectric coupling in the E-type commensurate perovskite manganese oxide compounds \cite{Kenzelmann2009}. For these materials a large ferroelectric polarization was predicted \cite{Sergienko2006,Picozzi2007} and recently found in TmMnO$_3$ \cite{Kenzelmann2009}.\\
Our theoretical description of the magnetoelectric effects in MnWO$_4$ differs in important aspects from the theoretical approach to this compound proposed by Harris \cite{Harris}. We determine from pure symmetry considerations the irreducible degrees of freedom (co-representations) involved in the observed sequence of phases, the corresponding order-parameter symmetries and the form of the transition free-energies and related magnetic, dielectric and coupling contributions, which allow describing the observed magnetoelectric effects. At last we take into account the actual positions of the Mn ions in order to establish the connection existing between our phenomenological order-parameters and the magnetic spins or electric dipoles. Harris follows an opposite procedure, starting from the actual magnetic structure which permits construction of allowed spin functions. These functions are then related to the order-parameter components, the transition free-energy and coupling to the polarization being deduced from semi-empirical considerations. The advantage of our approach is that it provides the full set of stable states allowed by the order-parameter symmetries (one of which being stabilized under high magnetic field), a detailed topology of the corresponding phase diagram, a faithful description of the critical behaviour, including the specific pseudo-proper (not improper) character of the induced polarization \cite{Toledano2009}, which has been overlooked by Harris. It also yields a precise explanation of the various magnetoelectric effects observed in MnWO$_4$ (not described by Harris) and an explicit determination of the different types of interactions contributing to the polarization.

\ack We acknowledge support from the Austrian FWF (P19284-N20) and the University of Vienna through the Focus Research Area \textit{Materials Science}.

\end{document}